\PassOptionsToPackage{table}{xcolor} 

\documentclass{acmart}

\usepackage[export]{adjustbox}
\usepackage[inline]{enumitem}
\usepackage{listings}
\usepackage[referable]{threeparttablex}
\usepackage{xspace}

\setcopyright{acmcopyright}
\copyrightyear{2023}
\acmYear{2023}
\acmDOI{XXXXXXX.XXXXXXX}

\begin{document}

\title{Application-Level Validation of Accelerator Designs Using a Formal Software/Hardware Interface}

\author{Bo-Yuan Huang}
\orcid{0000-0001-7069-4069}
\affiliation{
  \institution{Intel Corporation}
  \city{Santa Clara}
  \state{CA}
  \country{USA}
}
\authornote{Both authors contributed equally to this research, whilst at Princeton University and University of Washington, respectively.}

\author{Steven Lyubomirsky}
\orcid{0009-0003-6747-7014}
\affiliation{
  \institution{OctoML}
  \city{Seattle}
  \state{WA}
  \country{USA}
}
\authornotemark[1]

\author{Yi Li}
\orcid{0009-0000-4837-2282}
\author{Mike He}
\orcid{0009-0002-0843-8413}
\affiliation{
  \institution{Princeton University}
  \city{Princeton}
  \state{NJ}
  \country{USA}
}

\author{Gus Henry Smith}
\orcid{0000-0001-9754-233X}
\affiliation{
  \institution{University of Washington}
  \city{Seattle}
  \state{WA}
  \country{USA}
}

\author{Thierry Tambe}
\orcid{0000-0002-6411-9620}
\affiliation{
  \institution{Harvard University}
  \city{Cambridge}
  \state{MA}
  \country{USA}
}

\author{Akash Gaonkar}
\orcid{0000-0001-5565-2581}
\affiliation{
  \institution{Princeton University}
  \city{Princeton}
  \state{NJ}
  \country{USA}
}

\author{Vishal Canumalla}
\orcid{0009-0001-5418-1279}
\author{Andrew Cheung}
\orcid{0009-0006-0661-2640} 
\affiliation{
  \institution{University of Washington}
  \city{Seattle}
  \state{WA}
  \country{USA}
}

\author{Gu-Yeon Wei}
\orcid{0000-0001-5730-9904}
\affiliation{
  \institution{Harvard University}
  \city{Cambridge}
  \state{MA}
  \country{USA}
}

\author{Aarti Gupta}
\orcid{0000-0001-6676-9400} 
\affiliation{
  \institution{Princeton University}
  \city{Princeton}
  \state{NJ}
  \country{USA}
}

\author{Zachary Tatlock}
\orcid{0000-0002-4731-0124}
\affiliation{
  \institution{University of Washington}
  \city{Seattle}
  \state{WA}
  \country{USA}
}

\author{Sharad Malik}
\orcid{0000-0002-0837-5443}
\affiliation{
  \institution{Princeton University}
  \city{Princeton}
  \state{NJ}
  \country{USA}
}

\renewcommand{\shortauthors}{Huang and Lyubomirsky, et al.}

\ccsdesc[500]{Hardware~Application-specific VLSI designs}
\ccsdesc[300]{Hardware~Functional verification}
\ccsdesc[500]{Software and its engineering~Compilers}
\ccsdesc[300]{Computer systems organization~Heterogeneous (hybrid) systems}

\keywords{Accelerator, domain-specific language, compilation, validation, software/hardware interface}


\definecolor{halfgray}{gray}{0.5}
\definecolor{deepblue}{rgb}{0,0,0.5}
\definecolor{deepred}{rgb}{0.6,0,0}
\definecolor{deepgreen}{rgb}{0,0.5,0}
\definecolor{rered}{rgb}{0.88, 0.1, 0.26}
\definecolor{lightblue}{rgb}{0.9,0.95,1}
\definecolor{lightorange}{rgb}{1,0.97,0.90}
\definecolor{lightgray}{rgb}{0.96,0.96,0.96}
\lstset{
  basicstyle=\footnotesize\ttfamily,
  lineskip=0pt,
  basewidth=.54em,
  breaklines=false,
  showstringspaces=false,
  tabsize=2,
  keywordstyle=\bfseries\color{deepblue},
  emphstyle=\color{black},
  stringstyle=\color{deepred},
  commentstyle=\color{deepgreen},
  numbers=left,
  numberstyle=\scriptsize\color{darkgray}\ttfamily,
  numbersep=12pt,
  frame=l,
  framesep=7pt, 
  framextopmargin=0,
  framexbottommargin=0,
  framerule=0.5pt,
  rulecolor=\color{halfgray},
  xleftmargin=20pt, 
  captionpos=b
}
\lstdefinestyle{mapping-style}
{
  language=c,
  numbers=none,
  framerule=0pt,
  framesep=7pt,
  xleftmargin=15pt
}

\newlist{inlinelist}{enumerate*}{1}
\setlist*[inlinelist,1]{%
  label=(\arabic*)
}
\newcommand{\norm}[1]{\left\lVert#1\right\rVert}
\newcommand{\re}[1]{#1}
\newcommand{\recheck}[1]{#1}
\newcommand{\TLA}{3LA\xspace}
\newcommand{\AppNum}{six\xspace}
\newcommand{\egg}{\texttt{egg}\xspace}
\newcommand{\instrInText}[1]{\texttt{\small #1}}
\newcommand{\varInText}[1]{$\texttt{#1}$}
\newcommand{\cmark}{{\color{deepgreen}$\boldsymbol\vee$}}
\newcommand{\xmark}{{\color{deepred}$\boldsymbol\times$}}
\newcommand{\diy}{{\color{deepblue}$\boldsymbol\thicksim$}}
\newcommand{\mapping}{IR-to-accelerator mapping\xspace}

\begin{abstract}
Ideally,
  accelerator development
  should be as easy as
  software development.
Several recent design languages/tools
  are working
  toward this goal,
  but actually testing early designs
  on real applications end-to-end 
  remains prohibitively difficult
  due to the costs of building specialized
  compiler and simulator support.
We propose a new first-in-class, mostly automated methodology 
termed ``\TLA'' 
to enable
  end-to-end testing of prototype accelerator designs
  on unmodified source applications.
A key contribution of \TLA is the use of a formal software/hardware interface that specifies an accelerator's
  operations and their semantics. Specifically, we leverage the Instruction-Level Abstraction (ILA) formal specification for accelerators that has been successfully used thus far for accelerator implementation verification. We show how the ILA for accelerators serves as a software/hardware interface, similar to the Instruction Set Architecture (ISA) for processors, that can be used for automated development of compilers and instruction-level simulators.
  Another key contribution of this work is to show how ILA-based accelerator semantics enables
  extending recent work on equality saturation
  to auto-generate basic compiler support
  for prototype accelerators in a technique we
  term ``flexible matching.''
By combining flexible matching with
  simulators auto-generated from ILA specifications,
  our approach enables end-to-end evaluation
  with modest engineering effort.
We detail several case studies
  of \TLA, which uncovered
  an unknown flaw in
  a recently published accelerator and
  facilitated its fix.

\end{abstract}

\maketitle

\section{Introduction}
\label{sec.intro}


Hardware specialization 
  is the main technique 
  for improving power-performance efficiency in emerging compute platforms.
By customizing 
  compute engines, memory hierarchies, and data representations~\cite{chan2014itrs,fang2019understanding,lai2021programming},
  hardware accelerators provide 
  efficient computation in various application domains 
  like artificial intelligence, image processing, and graph analysis~\cite{han2016eie,chen2016eyeriss,reagen2016minerva,zhang2016cambricon,hameed2010understanding,ham2016graphicionado}.
However,
  despite significant recent progress in 
  design languages and tools for
  custom accelerators~\cite{nigam2021calyx, lai2019heterocl},
  many difficulties remain in
  developing domain-specific accelerators.

A particularly challenging aspect of accelerator development
  is validating early design prototypes
  on real applications.
Such validation is critical,
  as errors can arise from
  some of the techniques used
  to achieve maximum power-performance efficiency
  in accelerators,
  such as the use of custom numeric representations
  or reformulated operators.
In domains like deep learning (DL), signal processing or graphics,
  an application-level result
  (like a DL-based classification)
  can remain within acceptable range
  even if the numerical results
  of individual operations change slightly,
  presenting an opportunity to trade numerical accuracy
  for efficiency.
However, these changes need to be 
  carefully validated 
  at the application level---%
  even small changes in numerical accuracy
  of individual operators
  have the potential to cascade throughout an application,
  making the application-level results unacceptable~\cite{zorn2021rounding}.
Early end-to-end application level validation is essential
  for avoiding expensive and complex late stage hardware design changes.

\subsection{Challenges and Goals for Application-level Validation}
Testing accelerators under development
  on complete applications
  requires two critical components: compiler support and application-level testing support.  
\begin{itemize}
  \item {\bf Custom compiler support:}
  An application 
    (likely written in a domain-specific language, or DSL)
    must be adapted to offload computations to an accelerator,
    which entails writing DSL compiler passes
    or manual modification of the source program.
  In common practice,
    invoking accelerator operations from software
    requires engineering effort,
    such as developing custom drivers to invoke
    accelerators via memory-mapped I/O (MMIO) interfaces.
  Such drivers are opaque to the compiler,
    difficult to debug,
    and often rely on low-level architectural details.
  The compilation tasks would be simplified 
    through \emph{greater automation} in:
  \begin{inlinelist}
      \item identifying acceleration offload opportunities in the application, and 
      \item generating the low-level code that invokes the requisite accelerator operations.
  \end{inlinelist}

  \item {\bf Application-level testing support:}
  This goal poses several difficulties with existing techniques. 
  Register-transfer level (RTL) designs (and thus RTL simulation) are not available in the early stages when the proposed end-to-end-testing is most useful. Even when prototype RTL designs are available, RTL simulation is only practically feasible for individual operations, 
  being too slow for full applications.
  FPGA-based emulation requires significant engineering effort and is typically not done until late design stages. 
  Faster high-level simulation (e.g., using SystemC) is feasible,
  but requires manually writing detailed simulation models and
  verifying later that they are sound with respect to the RTL implementation.
  The ideal for application-level testing 
  is to \emph{automatically generate a sound high-level simulation model} for the accelerator that can be co-simulated with an application.
\end{itemize}

\begin{figure}[!ht]
  \begin{minipage}[h]{0.52\textwidth}
    \vspace{-5\fboxsep}
    \caption{
    \textbf{Gap in end-to-end evaluation of accelerators for neural network applications:} Our survey of $79$ papers in recent conferences (ISCA, MICRO, VLSI, and ISSCC in 2021 and ICCAD, DAC in 2020) that introduced new DL accelerator designs/methodologies, comparing how the accelerators were evaluated. Only 41\% of the works reported end-to-end evaluation on non-synthetic applications, of which 68\% (28\% of the total) were from industrial teams.
    }
    \label{fig.acc-survey}
  \end{minipage}\hfill
  \begin{minipage}[h]{0.43\textwidth}
    \includegraphics[width=\textwidth, right]{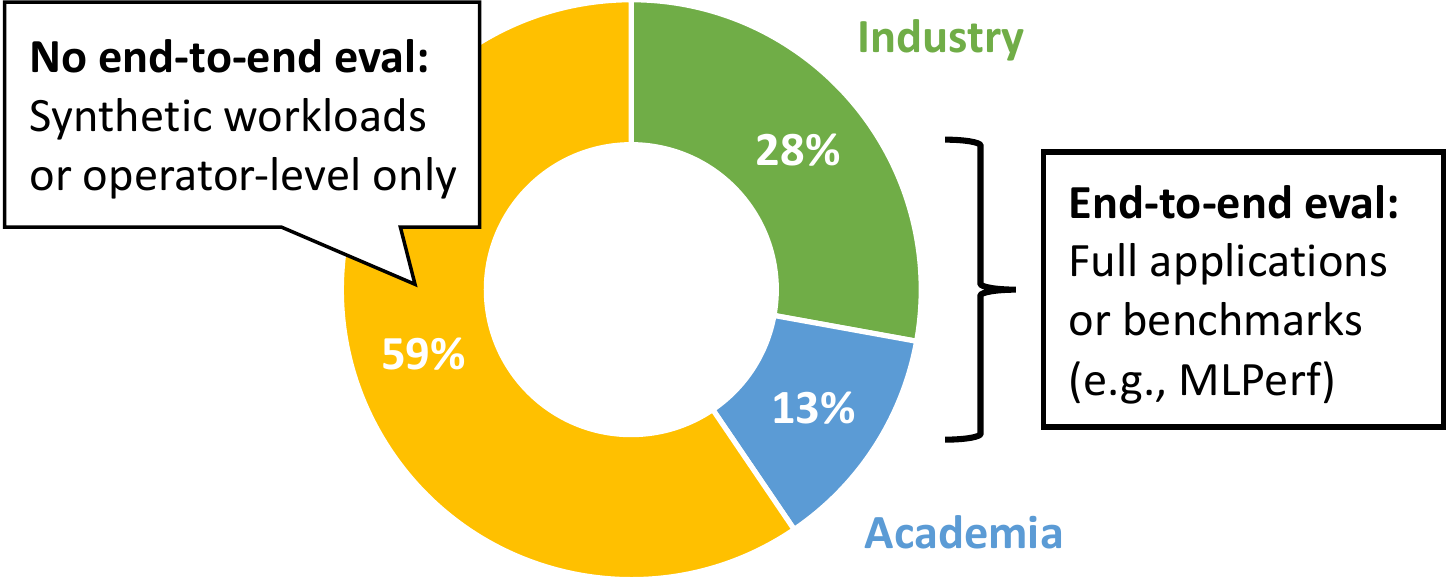}
  \end{minipage}
  \Description{}
\end{figure}

Note that the support components outlined above are specialized to a particular accelerator, and need to be updated every time an accelerator design is modified.
In current practice, large industrial teams 
  invest substantial resources to develop bespoke infrastructure~\cite{jouppi2017datacenter, jouppi2020tpu},
  while smaller teams often
  do not pursue end-to-end evaluation,
  as illustrated by our literature survey in Fig.~\ref{fig.acc-survey}. 

\subsection{Novel Contributions of our {\TLA} Approach}

We present a \emph{first-in-class} methodology 
  that supports
  end-to-end evaluation of accelerators
  {\it on unmodified full applications},
  which includes the ability to compile to
  and run simulations
  of 
  accelerator designs still in flux.
As a practical capability,
  \textit{
  this provides hardware designers with a feedback loop 
  similar to that of software debugging and testing.
  }

Our methodology,
  termed ``\TLA,'' 
  aims to reduce the manual engineering required for this feedback loop by effectively treating accelerator operations as extensions of processor instructions. 
  A novel contribution of \TLA is the use of a \emph{formal software/hardware (SW/HW) interface} that specifies an accelerator's
  operations and their semantics. Specifically, we leverage the Instruction-Level Abstraction (ILA), a formal specification for accelerators, that has been successfully used thus far for accelerator implementation verification~\cite{huang2018instruction} but not for compilation. In this work, we show how the ILA for accelerators serves as a SW/HW interface, similar to the Instruction Set Architecture (ISA) for processors, effectively serving as a ``single source of truth'' to drive various 
  tasks required for compilation and end-to-end application testing. (The same high-level ILA model can then be used in a late-stage ILA/RTL validation step using existing techniques.) While the ISA has wide applications in computer architecture/compilers, there is no existing framework that uses an ISA-like formal SW/HW interface for accelerators: {\TLA} provides \emph{an existence proof} that this is feasible both conceptually and as a practical framework. 
  
  
Our work makes the following novel contributions:
\begin{itemize}

  \item Use of a formal SW/HW interface for the accelerator: We use the ILA accelerator specification 
  to automate key tasks required for compilation and instruction-level simulation (\S\ref{sec:ila-compiling}). 
  Thus far, the ILA had only been used for accelerator implementation and firmware verification.
  
  \item Design of  ``flexible matching'' (\S\ref{sec.method.flexible}): This new semantics-guided term rewriting technique specialized to accelerators adds 
  custom rewrite rules for accelerators (\S\ref{sec:fragmentmapping}), and uses them in combination with  generic compiler intermediate representation (IR) rewrites. 
  This allows identifying, for the first time, semantically equivalent accelerator operations \emph{even without a direct syntactic match} 
  and significantly automates sophisticated operation offloading to accelerators without manually rewriting applications. 
  


  \item {\TLA} methodology and prototype: Combining the above techniques to achieve end-to-end mapping of unmodified applications to accelerators is another contribution. No existing tool (e.g., MLIR/CIRCT, PyMTL; see  \S\ref{sec:comparison}) 
  has attempted, much less achieved, the capabilities offered by the {\TLA} prototype (\S\ref{sec.prototype}) at such a level of automation. 
  Our evaluation (\S\ref{sec.eval}) demonstrates 
  automatic identification of multiple acceleration opportunities in off-the-shelf DL models imported from publicly available implementations and benchmarks. 
  We evaluated these models end-to-end in simulation for three different accelerators; this was the \textit{first time} that full applications were evaluated for two of the accelerators, and the tests exposed a flaw in one design related to numerical representations, which the developers were able to correct.

      
      

  \end{itemize}


The {\TLA} methodology requires two main inputs from the user:
\begin{inlinelist}
  \item ILA models: a formal ILA specification for accelerator operations (which can be reused for separate RTL verification), and
  \item \mapping rules: rewrite rules from the compiler IR to the accelerator operations (``mappings,'' for short).
\end{inlinelist}
Note that both of these are
  \textit{one-time efforts} per accelerator. 
Furthermore,
  this approach 
  greatly lowers the effort after hardware
  design revisions,
  as it requires modifying only the accelerator operation specifications and rewrite rules, if necessary,
  and obviates the need for
  certain additional work, like updating the high-level simulators, which are generated automatically in {\TLA}.

\subsection{Comparison with Existing Approaches and Tools}
\label{sec:comparison}

Although there have been many efforts in compiler flows to support accelerators~\cite{bahr2020creating,truong2020fault,lai2019heterocl,chen2021byoc,ragan2013halide,AtlPopl22,chen2018tvm,moreau2019hardware,lattner2021mlir,ExoPldi22}, none of them provides automated support for end-to-end testing of unmodified applications at the same level as {\TLA}. 
We start by providing a high-level comparison summarized in  
Table~\ref{figure.methodology} and provide a detailed comparison with specific tools at the end.

\subsubsection{Task-based Comparison}

Existing approaches use different techniques for three critical tasks:
accelerator operation selection, code generation, and software-hardware co-simulation. 
We compare them against {\TLA} for each task.

\paragraph*{Task 1: Accelerator Operation Selection} 
A common practice is for the software developer to manually insert API calls for accelerator invocations, or to use syntactic patterns to identify possible matches (e.g., using BYOC~\cite{chen2021byoc}). 
Bespoke compilation efforts, possibly built on top of tools like BYOC and frameworks like MLIR~\cite{lattner2021mlir} or Exo~\cite{ExoPldi22},
make sophisticated compilation passes to identify operations for offloading, 
but require compiler expertise. 
In contrast, \TLA overcomes the limitations of purely syntactic matching to find \emph{semantically-equivalent matches} in an automated way, without requiring the expertise and cost of bespoke compiler infrastructure.

\paragraph*{Task 2: Code Generation} 
This task involves emitting the actual instructions, i.e., the MMIO loads/stores, from the application program to invoke accelerator operations. 
A common practice is to emit MMIO code in implementations of the API calls that invoke accelerator operations, often referred to as the ``device driver'' for the accelerator. However, these API calls are opaque, in that a compiler has no built-in knowledge of the semantics of the accelerator operations or the MMIO code.  
Alternatively, in bespoke compilation efforts, this knowledge is built into the compiler but requires significant expertise and does not use a formal hardware semantics.
In contrast, in \TLA this code generation task is trivial following the operation selection step: each ILA instruction in the \mapping 
corresponds one-to-one with an MMIO instruction and is replaced accordingly. Moreover, the compiler has complete knowledge of their semantics via the formal SW/HW interface.


\paragraph*{Task 3: Software-Hardware Co-Simulation} 
Co-simulation is needed to validate the results of the computation being done by the accelerator offloads (the hardware) and the host processor (the software) for application-level testing. 

A common practice is to use frameworks like QEMU~\cite{bellard2005qemu}, which integrate RTL simulation calls (e.g., via Verilator~\cite{verilator}) with host processor execution. However, this is too slow for full-application testing.
Higher-level system models in languages like SystemC~\cite{SystemC} provide faster simulation but require significant effort for creating simulation models, and these models are difficult to validate against the RTL design.
In contrast, the ILA model in {\TLA} supports \emph{automatic generation of instruction-level simulation models} using the ILAng toolchain~\cite{huang2019ilang}. The ILA also allows separate verification against the RTL implementation, thereby ensuring soundness between the high-level simulation and the RTL implementation. 

\begin{table}[!ht]
\centering
\caption{
\textbf{Comparison of the \TLA methodology against existing approaches for three critical tasks.} 
Approaches discussed: 
TVM with BYOC~\cite{chen2021byoc}, Glenside~\cite{smith2021pure}, 
MLIR with various dialects~\cite{lattner2021mlir}, 
Halide with various extensions~\cite{liu2021-halide-stencils, vocke2017halide-im-dsps, pu2017-halide-stencils, li2020-halide-stencils-sysarrays, stratton2020-halide-dsp, gu2020-halide-im-dsp, schellekens2020-halide-stencils-auto, korhonen2015-halide-rapidcustomization}, 
Exo~\cite{ExoPldi22}, 
Verilator~\cite{verilator} with QEMU~\cite{bellard2005qemu}, SystemC~\cite{SystemC} with QEMU, PyMTL~\cite{batten2018pymtl}, and Catapult HLS~\cite{siemens-catapulthls}. 
}
\label{figure.methodology}
\begin{small}
\begin{tabular}{|lll|}

\hline
\textbf{Approach} & \textbf{Pros and Cons} & \textbf{Related Works} \\ \hline

\multicolumn{3}{|l|}{\textbf{Task 1: Accelerator operation selection}} \\
Manual selection & \cellcolor[HTML]{E9CECE}Simple, but tedious and error-prone & common practice \\
Syntactic pattern matching & \cellcolor[HTML]{E9CECE}Simple, but may miss accelerator offloads & BYOC, MLIR \\
Custom flow & \cellcolor[HTML]{F8E9AE}Flexible, but high effort to design (e.g., schedules of rewrites) & BYOC, MLIR, Halide, Exo \\
\TLA: ILA and mapping rules & \cellcolor[HTML]{DDEFDE}One-time effort (ILA + mapping rules) for Task 1-3  & Glenside (rewrite rules) \\ \hline

\multicolumn{3}{|l|}{\textbf{Task 2: Code generation}} \\

High-level API & \cellcolor[HTML]{E9CECE}No formal SW/HW interface, error-prone & common practice, BYOC \\
Bespoke codegen & \cellcolor[HTML]{E9CECE}No formal SW/HW interface, error-prone, high effort & BYOC, MLIR, Halide, Exo \\
\TLA: auto-gen. MMIO code & \cellcolor[HTML]{DDEFDE}Formal SW/HW interface, verifiable against RTL & \\ \hline

\multicolumn{3}{|l|}{\textbf{Task 3: Software-hardware co-simulation}} \\

RTL simulation & \cellcolor[HTML]{E9CECE}Late-stage, very slow, operation-level only & Verilator with QEMU \\
High-level software model & \cellcolor[HTML]{F8E9AE}Early-stage, end-to-end, not validated w.r.t. RTL & SystemC, PyMTL, Catapult \\
\TLA: auto-gen. ILA simulator & \cellcolor[HTML]{DDEFDE}Early-stage, end-to-end, validated w.r.t. RTL & \\ \hline
\end{tabular}
\end{small}
\end{table}

\subsubsection{Detailed comparison with closely related tools}
We discuss details of some specific tools.

\paragraph{MLIR~\cite{lattner2021mlir}} 
MLIR is a framework for building 
compiler IRs (as ``dialects'') in a structured, reusable manner.
Some MLIR dialects address tasks related to hardware design; these include CIRCT, which supports high-level synthesis (HLS) and hardware simulation but \emph{not compilation of applications to accelerators}. To the best of our knowledge, there is no SW/HW interface in MLIR that enables compiling applications to accelerators via CIRCT. Other dialects are intended to interface with specific accelerators (e.g., the TPU) and deep learning frameworks (e.g., ONNX).
However, compilation using these dialects still entails mapping between IRs at different granularities and other challenges, which are addressed by \TLA.

\paragraph{HLS tools (e.g., Catapult~\cite{siemens-catapulthls}) and PyMTL~\cite{batten2018pymtl}} 
These tools/frameworks allow for describing hardware designs with a high-level software-like interface, which is useful for designing accelerators and high-level hardware simulation. However, the hardware design specifications in these frameworks do not address accelerator operation selection and code generation during compilation---two of the three tasks in Table~\ref{figure.methodology}, which 
are significantly automated by {\TLA}. 

\paragraph{Exo~\cite{ExoPldi22}} 
Exo also addresses the problem of compiling applications to accelerators. It uses a notion similar to high-level instructions for interfacing with accelerators, but, unlike \TLA, relies on \emph{manually specified sequences of rewrites} and other bespoke compiler passes, to expose instances of those instructions and compile them to the devices using \emph{exact syntactic matching}. 
\TLA introduces flexible matching to \emph{automatically discover} accelerator offload opportunities in the applications, given a few rules relating the behavior of accelerator operations to the compiler IR (as in Exo). 
Additionally, Exo does not provide a \emph{formal SW/HW interface} or any means to verify its accelerator instructions against the RTL implementation.

\subsection{Paper organization}
We first provide the background (\S\ref{sec.background}) on ILA~\cite{huang2018instruction} and 
equality saturation~\cite{joshi2002denali, tate2011equality}.
The techniques in {\TLA} are described in detail next (\S\ref{sec.method}), followed by a description of the
{\TLA} prototype\footnote{Our prototype, benchmarks, and evaluation infrastructure will be open-sourced under a permissive license.} (\S\ref{sec.prototype}). 
We present detailed evaluation using our prototype (\S\ref{sec.eval}), and end with a discussion on more broadly related work (\S\ref{sec.related}) and conclusions (\S\ref{sec.conclusion}).
\section{Background}
\label{sec.background}


\subsection{ILA Software/Hardware Interface Specification}
\label{sec:ila}
The ILA is an ISA-like formal model 
  for specifying the functional behavior of accelerators.
It generalizes the ISA to accelerators, 
  where each 
  instruction of an accelerator ILA corresponds 
  to a command at the accelerator interface, i.e., an MMIO load or store from a host processor.
Like processor ISAs, 
  the ILA captures a formal semantics of the accelerator behavior, by specifying how each instruction 
  reads or updates software-visible (viz., architectural) state variables in the accelerator,
  while abstracting out implementation details. 
  
Fig.~\ref{fig.ila-example} shows an example
  ILA specification for one of the instructions of FlexASR~\cite{tambe20219} (one of the three accelerators used in our evaluation studies). The ILA models are written in ILAng, a DSL embedded in C++. The figure caption points out the per-instruction modular specification, where each instruction is specified by defining its decode condition (i.e., when the instruction is triggered) and state update functions (i.e., how it updates
  the architectural state variables). 

Thus far, the ILA has been used only for accelerator implementation verification and co-verification of firmware~\cite{huang2018formal,huang2018instruction}.
In this work, we use ILA as a formal SW/HW interface that drives the key tasks in compilation and application-level testing for accelerators.

\begin{figure}[!ht]
  \begin{minipage}[h]{0.31\textwidth}
    \vspace{-3\fboxsep}
    \caption{
    \textbf{ILA model (snippet) for FlexASR accelerator.}
    Lines 3-11 define the inputs and architectural state variables.
    Lines 14-22 show an example ILA instruction 
    ``\texttt{\small pe\_0\_cfg\_mngr}.'' 
    Its decode condition (lines 16-17) specifies that this instruction is triggered when there is a write command to the 
    processing element's (PE) management configuration register.
    Its state update functions (lines 19-22) specify that this instruction stores arguments from the interface inputs into the corresponding configuration registers. The state update functions of instructions such as for linear layer (elided here) 
    encode their operational semantics.
    This snippet highlights:
    (1) similarity of ILA with ISA, and
    (2) the formal semantics based on how each instruction reads/writes the architectural state variables.
    }
    \label{fig.ila-example}
  \end{minipage}\hfill
  \begin{minipage}[h]{0.66\textwidth}
    \begin{lstlisting}[language=c]
auto m = ilang::Ila("flexasr-ila");
// declare inputs at the interface
auto wr = m.NewBvInput("top_if_wr", TOP_IF_WR_BITS);
auto rd = m.NewBvInput("top_if_rd", TOP_IF_RD_BITS);
auto addr = m.NewBvInput("top_addr_in", TOP_ADDR_IN_BITS);
auto data = m.NewBvInput("top_data_in", TOP_DATA_IN_BITS);
// declare architectural states
m.NewBvState("pe_0_is_valid", PE_VALID_BITS);
m.NewBvState("pe_0_is_bias", PE_IS_BIAS_BITS);
m.NewMemState("gb_large_buffer", TOP_ADDR_IN_BITS, TOP_DATA_IN_BITS);
// ... (some code)
// ILA instruction for configuring pe_cfg_mngr
auto instr = m.NewInstr("pe_0_cfg_mngr");
// define decode condition for this instruction
auto is_write = (wr == 1) & (rd == 0);
instr.SetDecode(is_write & (addr == PE_0_CFG_MNGR_ADDR));
// define state update functions for this instruction
auto is_valid = ilang::SelectBit(data, PE_IS_VALID_BIT_IDX);
instr.SetUpdate(m.state("pe_0_is_valid"), is_valid);
auto is_bias = ilang::SelectBit(data, PE_IS_BIAS_BIT_IDX);
instr.SetUpdate(m.state("pe_0_is_bias"), is_bias);
// ... (more code)
    \end{lstlisting}
  \end{minipage}
\Description{}
\end{figure}

\subsection{Term Rewriting with Equality Saturation}
\label{sec.rewrite}

Term rewriting is a well-known technique 
  for program transformations,
  with some compiler optimizations being implemented
  as term-rewriting systems~\cite{
    dershowitz1993taste,
    baader1999term,
    blindell2016instruction,
    regis-pact22}.
Given a set of syntactic rewrite rules ($\ell \rightarrow r$) that also preserve semantic equality, 
  a term-rewriting system 
  rewrites instances of pattern $\ell$ 
  in the input program with semantically equivalent pattern $r$ where applicable.
  
In traditional term rewriting,
  applying one rewrite rule
  may prevent
  using other, potentially profitable, rewrite rules;
  this is referred to as the phase-ordering problem~\cite{whitfield1997approach}.
Equality saturation avoids
  phase-ordering issues 
  by searching over many equivalent rewritings of the same program~\cite{tate2011equality,joshi2002denali}.
Given an input program $p$, 
  equality saturation repeatedly applies 
  the given rewrite rules 
  to explore all equivalent ways to express $p$
using an \textit{e-graph} data structure
  to efficiently represent an exponentially large set of equivalent program expressions~\cite{nelson1980fast,nieuwenhuis2005proof}.
Upon reaching a fixed point, 
  i.e., when no application of any rewrite rule can introduce a new program expression,
  or upon hitting a predetermined resource limit,
  the optimal rewritten program
  can be extracted from an e-graph
  according to a given cost function.

In {\TLA}, we extend equality saturation to support accelerators. 
Specifically, we
 create custom rewrite rules for accelerators (\S\ref{sec:fragmentmapping}), 
and specialize the cost function to maximize the number of accelerator offloads (\S\ref{sec.method.flexible}) or 
consider cost of data movement (\S\ref{sec:optimization}).
  Our prototype uses
  the \egg library~\cite{willsey2021egg} for its
  efficient implementation of equality saturation.
\section{{\TLA} Methodology}
\label{sec.method}

\begin{figure}
\centering
\includegraphics[width=.55\textwidth]{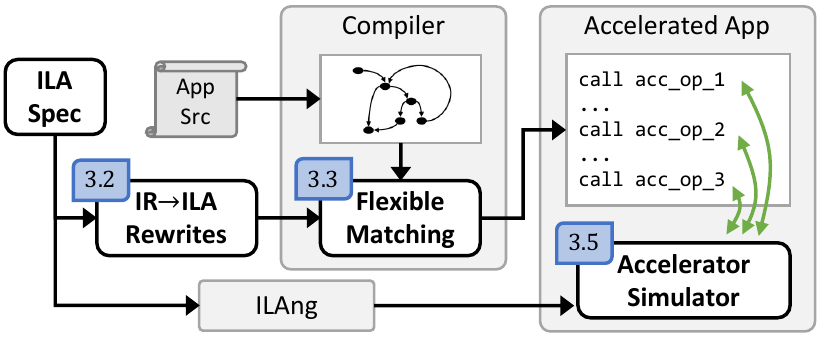}
\caption{
\textbf{\TLA methodology flow.}
\S\ref{sec:fragmentmapping}: the developer provides one 
\emph{\mapping}
  for each 
  accelerator operator,
\S\ref{sec.method.flexible}: a compiler extended with \emph{flexible matching}
  automatically maps unmodified source application fragments 
  to accelerator calls, 
\S\ref{sec:cosimulation}: the ILAng platform generates fast simulators, enabling end-to-end co-simulation.
These features enable a mostly automated workflow for end-to-end testing of prototype accelerator designs on unmodified applications. 
  }
\label{fig.abstract-flow}
\Description{}
\end{figure}

We now describe technical details of the {\TLA} methodology,
 along with illustrative examples
  that demonstrate its various components.
The important steps in the \TLA flow are shown in Fig.~\ref{fig.abstract-flow}, and described in the related subsections. 
Our examples and prototype (\S\ref{sec.prototype}) 
  pertain to deep learning,
  but 
  the {\TLA} techniques based on a formal SW/HW interface and flexible matching are general and can apply to other domains.

We consider the following two examples: 
\begin{inlinelist}
\item an LSTM word language model (LSTM-WLM), a text generation application 
  consisting of an LSTM recurrent neural network 
  (RNN) followed by a linear 
  layer~\cite{pt2020wlm, graves2014towards}, and
\item ResNet-20, a widely used image classification model
  featuring 2D convolutions 
  and residual connections~\cite{he2016deep}.
\end{inlinelist}
Since the LSTM RNN comprises most of the computation in 
  LSTM-WLM, it is desirable to accelerate this application 
  using FlexASR~\cite{tambe20219},
  a natural language processing accelerator
  that includes
  support for both LSTM RNNs and linear layers in hardware.
  Thus, FlexASR can also be used to accelerate the linear layers 
  in ResNet-20. In addition, ResNet-20 can also be accelerated with HLSCNN~\cite{whatmough201916nm},
  an accelerator for 2D convolutions exclusively;
  hence, the two accelerators can be used in concert. 

  However, compiling DL applications like these
  from a high-level DSL 
  (e.g., PyTorch for LSTM-WLM)
  to coarse-grained accelerators like FlexASR and HLSCNN
  poses several challenges highlighted below:
\begin{enumerate}
    \item \textbf{Specialized accelerator interfaces.} 
      These accelerators, 
        like many others,
        are invoked using MMIO instructions over AXI interfaces,
        to configure the accelerator's state and signal when to begin operations,
        thus requiring a thorough knowledge of 
        both the accelerator architecture and functions of the MMIO instructions. 
    \item \textbf{Granularity mismatch.}
      The compiler must relate
        the accelerator operations' coarse-grained semantics
        (e.g., an LSTM RNN) with the possibly fine-grained corresponding representations in the compiler IR.
    \item \textbf{Numerical representations.}
      Accelerators often use specialized numerical representations for improved performance or reduced hardware costs. For example, 
        FlexASR uses a custom format called AdaptivFloat~\cite{tambe2020algorithm}
        and HLSCNN uses a mixed 8/16-bit fixed-point representation.
        One must check that these data types
        do not cause inaccuracies at the full application
        level, 
        particularly when values are cast between data types.
\end{enumerate}
In the descriptions of various {\TLA} techniques in this section, we will emphasize how they address these challenges. 

\subsection{The ILA as a Formal SW/HW Interface in {\TLA}}
\label{sec:ila-compiling}
The first step in the {\TLA} methodology is to develop the ILA formal models for accelerators (\S\ref{sec:ila}). We follow the techniques proposed in prior work~\cite{huang2018instruction,huang2019ilang}, where each instruction of an accelerator ILA corresponds 
  to a command at the accelerator interface. Some instructions are simple instructions that configure the accelerator, while others may trigger complex operations, e.g., FlexASR's linear layer operator.
  
The ILA for accelerators serves as a formal software/hardware interface in {\TLA}, similar to the Instruction Set Architecture (ISA) for processors, and effectively drives the following tasks required for compilation and end-to-end application testing with accelerators.
\begin{itemize}
\item {\it Accelerator Operation Selection:} A formal instruction representation for accelerator operations enables adapting existing instruction selection techniques
  to identify acceleration opportunities and 
  output the ILA instructions that correspond to functionally equivalent parts of the compiler IR representation. 
\item {\it Code Generation:} The 
 ILA instructions correspond one-to-one 
 with the MMIO commands that operate the accelerator.
 Thus, the selected ILA instructions can be directly lowered
 to MMIO commands for invoking accelerator operations from the application code running on a host processor.
\item {\it SW/HW Co-Simulation}: The ILAng toolchain~\cite{huang2019ilang}
  can automatically generate
  a functional simulator
  given an ILA specification---%
  this can be done in the early design stages, 
  even without an RTL implementation.
  \end{itemize}
Further, note that a revision to the accelerator design can easily be tested
  simply by making changes to the comparatively high-level ILA specification.



The ILA model size is about 10-20\% of the size of the RTL implementation~\cite{huang2018instruction}. Although writing the ILA model can be a significant one-time effort, it carries benefits
beyond the context of the {\TLA} methodology.
The ILA specification has been used for checking the accelerator RTL implementation,
  both using formal verification~\cite{huang2018instruction} 
  and simulation-based validation~\cite{huang2019ilang}.
Thus, ILA-based accelerator simulations in {\TLA} can be made sound with respect to RTL. 

\subsection{Compiler IR-to-Accelerator Mapping}
\label{sec:fragmentmapping}


To support compilation to accelerators, 
  we require some means of mapping from the compiler IR  
  to the ILA instructions that specify the accelerator operations.
  This is accomplished by specifying
  an \emph{\mapping rule} (``mapping,'' in short). 
  In general, this is a many-to-many mapping, 
  i.e., where a program fragment with many instructions in the compiler IR are rewritten to a program fragment with many ILA instructions on the accelerator side.
This provides a general way to handle different granularities in compiler IR intrinsics (e.g., dot products and convolutions) and in accelerator operations (e.g., fine-grained operations in VTA~\cite{moreau2019hardware} and coarse-grained operations in HLSCNN, FlexASR). Furthermore, the ILA instructions in the mapping provide a verifiable abstraction of the hardware accelerator operation in terms of updates to software-visible architectural state. 

Writing the 
\mapping rule
  is a one-time effort per accelerator operation 
  and is reusable across applications.
Further, 
  the mapping rule 
  can be validated by comparing the results of the compiler IR fragment on the host device and the ILA-based simulation results. 
  

\textbf{Examples.} 
In our prototype,
  we use the Relay IR~\cite{roesch2019relay}
  in the TVM DL compiler stack~\cite{chen2018tvm}
  as the compiler IR.
TVM supports importing models from other DL frameworks
  by converting them into Relay,
  thus allowing our prototype to support these front-ends as well.
Further, it enables 
  leveraging the Bring Your Own Codegen (BYOC)~\cite{chen2021byoc} library for code generation (discussed later).
For LSTM-WLM on FlexASR,
  we provide 
  a mapping from an LSTM RNN
  (a large construct in Relay if ``unrolled'')
  to a short sequence of FlexASR ILA instructions, 
  and another mapping for a linear layer, which we illustrate in Fig.~\ref{fig.fragment-mapping}.
ResNet-20 also uses the same mapping for linear layer
  and a straightforward 
  mapping
  from a single 2D convolution operator 
  to a sequence of HLSCNN ILA instructions
  for performing a convolution.

\begin{figure}
  \centering
  \input{Floats/lst-flexasr-mapping}
  \caption{
    \textbf{An example \mapping for the FlexASR linear layer operation.} 
    The compiler IR fragment (a) is mapped to a sequence of FlexASR ILA instructions (b) that configure the accelerator states and trigger the computation.
    The ILA instructions correspond one-to-one to the accelerator's MMIO commands (c).
    This example illustrates how the ILA instructions are used in mapping rules and code generation.
  }
  \label{fig.fragment-mapping}
  \Description{}
\end{figure}

\subsection{Flexible Matching for Accelerator Operator Selection}
\label{sec.method.flexible}

Given compiler \mapping rules, 
  we can identify all potential 
  offloads to an accelerator 
  by finding portions of an application
  that match the given compiler IR fragments, syntactically or semantically.

\subsubsection{Difficulties due to syntactic matching}
Searching the application
  for exact syntactic matches
  for the given compiler IR fragments
  (referred to as ``exact matching'')
  is simple to implement 
  (e.g., this is done by the BYOC library in TVM~\cite{chen2021byoc}).
However, exact matching faces difficulties
  as there is often no canonical way
  to represent an operation,
  necessitating either the addition of more patterns
  or manual modifications to the input program
  to match the expected patterns.
Developing a canonicalization for each given IR may be possible,
  but would require careful design per IR and further effort to prove that the program transformations preserve the canonicalization~\cite{newcomb20-halide-verif}.
Application code can vary greatly in structure,
  particularly in the case of compiler IRs,
  which may be produced after several iterations
  of program transformations
  (as with TVM, its model importers
  may translate equivalent expressions
  from various frameworks
  into different Relay expressions).


\textbf{Examples.} 
In our LSTM-WLM, the compiler IR pattern for a linear layer is (as an S-expression~\cite{mccarthy1960sexp}):
\[ \texttt{(bias\_add (nn\_dense \%a \%b) \%c)}. \]
%
However, 
  in ResNet-20,
  which was imported from MxNet,
  linear layers are equivalently expressed as: 
\[ \texttt{(add (reshape (nn\_dense \%a \%b) \%s) \%c)} \]

when \instrInText{\%c} is a vector, for certain shapes \instrInText{\%s}.
The former pattern would fail to match it,
  thus missing an opportunity to invoke FlexASR's linear layer operation.

\subsubsection{Semantic matching via term rewriting} 
\label{sec:semantic}

 Rather than attempt to enumerate all semantically equivalent patterns 
 (a task that is tedious, error-prone, and likely to result in an incomplete enumeration), 
 or expect users to modify their application code to expose expected patterns (demanding knowledge of the model and patterns
  as well as engineering effort), 
   \TLA  aims to maximize the degree of automation
  by utilizing term-rewriting and equality saturation techniques to transform programs
  \emph{to expose the most  matching opportunities for accelerator operation selection}. We call this process ``flexible matching'', and describe how it is specialized for accelerators. 

\re{
Flexible matching uses two kinds of rewrite rules:
\begin{itemize}
\item Compiler IR rewrite rules: These are general-purpose rules, independent of the accelerator, and are reusable and composable for various applications. We have developed a general set in \TLA 
including rules for, e.g., merging/splitting tensors, commutativity, associativity, and identities for common operators. 

\item \mapping rules: These rewrite rules are accelerator-specific. Recall (\S\ref{sec:fragmentmapping}) that these mappings are many-to-many, providing a general way to handle different granularities in compiler IR intrinsics and accelerator operations.
When targeting new accelerators, accelerator designers are expected to provide these mappings. For our evaluation, we created these mappings for the operations supported by the accelerators. 
\end{itemize}

All rewrites in {\TLA} are polymorphic over tensor size, which requires specifying relationships between the input and output sizes for operations that merge, split, or broadcast over tensors. This also makes a given \mapping more general and provides support for applications using different block sizes, strides, etc., without changing any rules. 

One benefit of separating the two kinds of rewrite rules in flexible matching is that this allows the compiler IR rewrites to use purely functional IRs, without requiring bespoke compilation steps for state/effect analysis during those rewrites, while stateful effects are limited to mappings, where they are formally specified by ILA instructions. 
Another benefit is  
  that mappings for multiple accelerators
  can be \emph{simultaneously} included, 
  thereby searching over all opportunities 
  to invoke all available accelerators in concert.
}

In the extraction phase of equality saturation, the rewritten program optimizing the cost function is chosen. 
This provides flexibility in the criteria for selection among functionally equivalent candidates for accelerator offloads. In our evaluations where we focused on end-to-end functional testing, we used a simple cost function that maximizes the number of accelerator invocations. More sophisticated cost functions can incorporate information about performance or data movement costs, and thereby result in different offloads.

\textbf{Examples.}
In our prototype,
  we compile programs in Relay
  into another IR called Glenside~\cite{smith2021pure}, 
  which uses the \egg library~\cite{willsey2021egg}
  to implement equality saturation
  for tensor programs.
In addition to matching the compiler \mapping 
  by adding them as rewrite rules in Glenside,
  this approach also facilitates \emph{additional} matches
  through the inclusion of \textit{general-purpose rewrite rules}
  within Glenside,
  such as tensor shape transformations and algebraic manipulations of combinators.
These general-purpose rules
  allow the term-rewriting system to conclude
  that different variations of an expression
  (like the linear layer examples above)
  are, in fact, equivalent.
In our examples, using the rewrite rules in Glenside
  exposed acceleration opportunities in both
  the LSTM-WLM and ResNet-20 programs
  through specifying \textit{only a single}
  \mapping rule for each accelerator operator.

It was not clear \textit{a priori} whether flexible matching would be performant for accelerators with complex \mapping rules needed for available accelerator designs. Our evaluation results (\S\ref{sec.eval}) show that powerful compiler IR rewrites can be combined effectively with a few \mapping rewrites in flexible matching, which finds more matches than exact matching in reasonable time.

\subsection{\TLA support for additional optimizations}
\label{sec:optimization}
The following \TLA design features provide support for additional optimizations:
\begin{inlinelist}
    \item 
\mapping: 
Recall that 
our mappings are polymorphic over tensor size, i.e., we parameterize these mappings with arguments such as sizes of the input/output data.
On the accelerator side, the maximum sizes supported by a single accelerator offload are limited by the accelerator-controlled hardware resources (e.g., buffer sizes, memory layout, internal/external memory, etc.). 
    \item
Flexible matching: Our technique provides optimization capabilities by including performance or other criteria in the cost function, which is optimized for instruction selection. 
\end{inlinelist}

In ongoing \TLA work, these features have supported the design of a ``scheduler'' that decomposes applications with oversized layers to accelerator operations; i.e., it determines loop orders, tile sizes, etc., based on a specification of the accelerator-controlled memory/hardware resources. For our current set of accelerators, our scheduler generates a guaranteed optimal schedule with the least data movement between the host and accelerator for mapping single layers in an application. It includes several novel techniques to prune the large search space (beyond the scope of this paper). Importantly, it is fast enough (a few seconds in our experiments) to be used for cost estimation during flexible matching.

In future work, we plan to integrate this scheduler with flexible matching, thereby enabling optimization of data movement 
or for considering possible tradeoffs with other costs. 
Another optimization opportunity we have identified is in removing redundant intermediate data transfers in back-to-back offloads to accelerator operations. This can be done in a pass after flexible matching and before code generation.

\subsection{Co-Simulation for Application-Level Results}
\label{sec:cosimulation}

  After the acceleration operation selection is done and specific portions of the application are marked as offloaded to an accelerator, 
  we can co-simulate the results at the full-application level  
  rather than for only individual operators.
Namely, the portions of the applications that are not marked
  are directly executed on the host 
  (generally a CPU)
  and the marked portions converted into their corresponding ILA instruction sequences
  are 
  simulated
  via an ILAng-generated simulator.
Note that the ILAng-generated simulators faithfully simulate the custom numerics, either using semantics formally modeled in the ILA specification or by accepting trusted software libraries that implement custom numerical data types.

\textbf{Examples.} 
In \S\ref{sec.eval}, we examine the application of the \TLA methodology on several DL applications.
In the process,
  upon identifying a numerical accuracy issue with HLSCNN in ResNet-20 and MobileNet-V2,
  we rapidly explored revisions to the design
  by changing the ILA specification---%
  a much simpler task than modifying the RTL.
\section{\TLA Prototype Implementation}
\label{sec.prototype}

\begin{figure*}
  \centering
  \includegraphics[width=\linewidth]{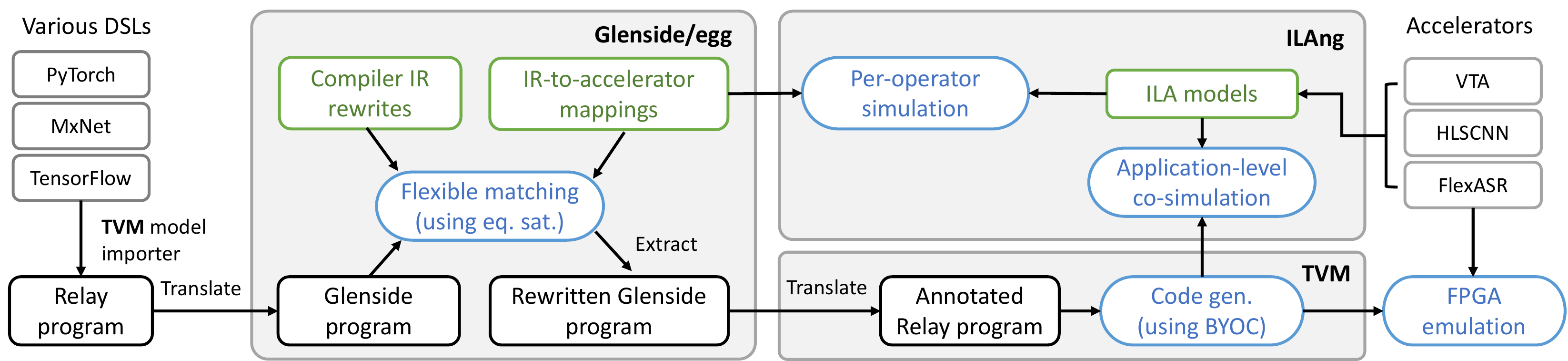}
  \caption{
  \textbf{Prototype implementation of the {\TLA} flow.}
  The green boxes represent additional inputs needed for the \TLA flow. The blue boxes represent the mostly automated capabilities added by the flow. The ILA specification model can also be used to formally verify the accelerator RTL implementation (as demonstrated in prior work~\cite{huang2018instruction}). The compiler IR rewrite rules and \mapping 
  rules can support other 
  compiler optimization and verification tasks.
  }
  \label{fig.prototype}
\Description{}
\end{figure*}

As a demonstration of the {\TLA} methodology, we have implemented an end-to-end compilation and simulation flow for DL applications by integrating with existing compiler frameworks and the ILAng platform~\cite{huang2019ilang}, as shown in Fig.~\ref{fig.prototype}.
Specifically, 
  our prototype is integrated into the TVM DL compiler and uses Relay
  as the representation for DL applications~\cite{chen2018tvm, roesch2019relay}.
We convert Relay programs into Glenside, and then perform flexible matching via equality saturation on tensor programs using \egg~\cite{smith2021pure,willsey2021egg}.
Finally, we use ILAng for co-simulating the compiled applications.

%


\paragraph{DSL Front-End}
TVM is a compiler framework for DL applications. 
We make use of TVM's model importer as the front-end for DSL programs.
The importer takes programs written in common DL DSLs
(e.g., ONNX~\cite{linux2019onnx}, PyTorch~\cite{paszke2019pytorch}, and TensorFlow~\cite{abadi2016tensorflow}) and translates them into Relay.


\paragraph{Flexible Matching}
As described earlier (\S\ref{sec.method.flexible}),
  we implement flexible matching
  by translating input
  programs from Relay
  into Glenside.
Given both compiler IR rewrites and 
{\mapping}s 
via Glenside, \egg explores the space of acceleration opportunities using equality saturation.

\paragraph{Code Generation}
Once flexible matching completes, the extracted rewritten program is translated back to Relay 
where accelerator 
operations 
are specially annotated. 
%
In our prototype, we use TVM's BYOC library~\cite{chen2021byoc}  to implement code generation (i.e., MMIO instructions and data movement code) for 
these accelerator operations. 
\section{Evaluation}
\label{sec.eval}

In this section, 
  we evaluate our prototype 
  for end-to-end testing with \AppNum applications and three accelerator designs. We focus especially on
\begin{inlinelist}
\item automated identification of acceleration opportunities, and 
\item application-level validation using automated co-simulation.   
\end{inlinelist}
%
Note that other tools provide very little automated support (if any)
for these two capabilities, thus precluding any head-to-head comparisons. 
We also report on operator-level evaluation (accuracy and performance) and FPGA-based deployment.


\subsection{Target Accelerators}
\label{sec.eval-acc}

We added support for three DL accelerators that provide hardware operators at different levels of granularity:
\begin{enumerate}[leftmargin=*]

\item \textbf{FlexASR} is an accelerator for speech and natural language processing (NLP) tasks that supports various RNNs~\cite{tambe20219}.
It uses a custom numeric data type called \textit{AdaptivFloat} for boosting the accuracy of quantized computations~\cite{tambe2020algorithm}.
%
%
%


\item \textbf{HLSCNN} is an accelerator optimized for 2D convolutions~\cite{whatmough201916nm}.
%
%
It operates on mixed 8/16-bit fixed point data
(8 bits for storing weights and 16 bits for computations).
%

\item \textbf{VTA} is a parameterizable accelerator for tensor operations featuring a processor-like design~\cite{moreau2019hardware}.
%
It supports element-wise arithmetic operations as well as generalized matrix multiplication,
operating on 8-bit integer data.
%

\end{enumerate}
%
For each accelerator, we defined an ILA model and a set of \mapping rules.
The ILA models for FlexASR, HLSCNN, and VTA are approximately 5600, 1600, and 2100 lines of ILAng code (C++), respectively. 
The high-level synthesis (HLS) implementations
  of the accelerators are about 9300 (SystemC), 5100 (SystemC), and 6900 (Chisel) LoC, 
  respectively;
  the ILA specifications are thus of modest size,
  compared even to the relatively compact HLS implementations. 
For each \mapping rule, we represent the compiler side in Glenside IR, and the accelerator side 
as a program composed of ILA instructions (in a Python-embedded DSL). 
The total size of mapping rules (both the compiler and accelerator sides) for FlexASR (5 mappings), HLSCNN (1 mapping), and VTA (1 mapping) was 186, 22, and 49 LoC, respectively.
Recall that these mappings are polymorphic over tensor size on both sides, leading to general and compact representations. 
Additionally, the BYOC-based code generators and runtimes for these accelerators are approximately 450, 300, and 900 LoC of C++, 
respectively. 
These indicate the implementation of the code generation module in our prototype, as well as reusable utilities for data movement, handling custom numerics, and emitting the low-level MMIO code for each selected accelerator offload for end-to-end simulation of the application.

%
%

\subsection{Target Applications}
\label{sec.eval-app}



We considered \AppNum DL applications corresponding to common neural network models for language and vision tasks that contain operators supported by the three target accelerators.
We selected applications with reasonable size for human inspection and in-depth analysis.
%
%
\begin{enumerate}[leftmargin=*]


\item \textbf{EfficientNet} is a recent convolutional neural network (CNN) designed for image classification~\cite{tan2019efficientnet}. 
It has convolutions that are supported by VTA and HLSCNN.



\item \textbf{LSTM-WLM} is a text generation application~\cite{pt2020wlm} implemented using an LSTM recurrent neural network architecture~\cite{graves2014towards}. The LSTM layer in this model is supported by FlexASR.


\item \textbf{MobileNet-V2} is a common CNN designed for mobile applications~\cite{howard2017mobilenets, sandler2019mobilenetv2}. We chose MobileNet-V2 due to its wide use, especially on embedded devices.

\item \textbf{ResMLP} is a recent residual network for image classification, comprised only of multi-layer perceptrons ~\cite{touvron2021resmlp}. Its linear layers could be accelerated by VTA and FlexASR.


\item \textbf{Transformer} is an NLP model comprised primarily of attention mechanisms~\cite{vaswani2017attention}. 
We chose Transformer as a representative of recent popular NLP models. 

\item \textbf{ResNet} is a popular CNN designed for image classification~\cite{he2016deep}. 
Besides ResNet-20, which we use in most of the evaluation, in \S\ref{sec.compilation-stats}, we additionally compare various implementations of ResNet-50 from MLPerf~\cite{mlperf} for its availability of different reference implementations.


\end{enumerate}
%
All applications were mapped to accelerators \emph{without any manual modifications}.
\begin{table*}
  \centering
  \caption{
  \textbf{End-to-end compilation statistics.}
  The total number of Relay operators (row 3) is given as a proxy
  for program complexity.
  In rows 4-6, we include rewrites for only one accelerator at a time; we do not offload to multiple accelerators at once like in \S\ref{sec.end-to-end}. Flexible matching identifies significantly more offloads than exact matching.
  Abbreviations: MN: MobileNet, Trans.: Transformer, and TF: TensorFlow.
  }
  \label{tab.compilation}
  \begin{small}
  \begin{tabular}{|clrrrrrr||rrr|}
  \hline
  \multicolumn{11}{|c|}{Application Statistics} \\ \hline
  \multicolumn{1}{|c|}{1} &
    \multicolumn{1}{|l|}{Application} &
    \multicolumn{1}{c|}{EfficientNet} &
    \multicolumn{1}{c|}{LSTM-WLM} &
    \multicolumn{1}{c|}{MN-V2} &
    \multicolumn{1}{c|}{ResMLP} &
    \multicolumn{1}{c|}{Trans.} &
    \multicolumn{1}{c||}{ResNet-20} &
    \multicolumn{3}{c|}{ResNet-50}
    \\ \hline
  \multicolumn{1}{|c|}{2} &
    \multicolumn{1}{|l|}{Source DSL} &
    \multicolumn{1}{c|}{MxNet} &
    \multicolumn{1}{c|}{PyTorch} &
    \multicolumn{1}{c|}{PyTorch} &
    \multicolumn{1}{c|}{PyTorch} &
    \multicolumn{1}{c|}{PyTorch} & 
    \multicolumn{1}{c||}{MxNet} &
    \multicolumn{1}{c|}{PyTorch} &
    \multicolumn{1}{c|}{ONNX} &
    \multicolumn{1}{c|}{TF}
    \\ \hline
  \multicolumn{1}{|c|}{3} &
    \multicolumn{1}{|c|}{\#Relay Ops} &
    \multicolumn{1}{c|}{232} &
    \multicolumn{1}{c|}{578} &
    \multicolumn{1}{c|}{757} &
    \multicolumn{1}{c|}{343} &
    \multicolumn{1}{c|}{872} &
    \multicolumn{1}{c||}{494} &
    \multicolumn{1}{c|}{709} &
    \multicolumn{1}{c|}{194} &
    \multicolumn{1}{c|}{609}
    \\ \hline \hline
  \multicolumn{11}{|c|}{Number of Static Accelerator Offloads Identified Using Exact Matching/Flexible Matching} \\ \hline
  \multicolumn{1}{|c|}{4} &
    \multicolumn{1}{|l|}{FlexASR} &
    \multicolumn{1}{r|}{0/35} &
    \multicolumn{1}{r|}{1/1} &
    \multicolumn{1}{r|}{0/41} &
    \multicolumn{1}{r|}{0/38} &
    \multicolumn{1}{r|}{0/66} &
    \multicolumn{1}{r||}{2/22} &
    \multicolumn{1}{r|}{0/54} &
    \multicolumn{1}{r|}{0/54} &
    \multicolumn{1}{r|}{0/54}
    \\ \hline 
  \multicolumn{1}{|c|}{5} &
    \multicolumn{1}{|l|}{HLSCNN} &
    \multicolumn{1}{r|}{35/35} &
    \multicolumn{1}{r|}{0/0} &
    \multicolumn{1}{r|}{40/40} &
    \multicolumn{1}{r|}{0/0} &
    \multicolumn{1}{r|}{0/0} &
    \multicolumn{1}{r||}{21/21} &
    \multicolumn{1}{r|}{53/53} &
    \multicolumn{1}{r|}{53/53} &
    \multicolumn{1}{r|}{0/53}
    \\ \hline 
  \multicolumn{1}{|c|}{6} &
    \multicolumn{1}{|l|}{VTA} &
    \multicolumn{1}{r|}{0/35} &
    \multicolumn{1}{r|}{36/36} &
    \multicolumn{1}{r|}{1/41} &
    \multicolumn{1}{r|}{38/38} &
    \multicolumn{1}{r|}{66/66} &
    \multicolumn{1}{r||}{0/22} &
    \multicolumn{1}{r|}{0/24} &
    \multicolumn{1}{r|}{0/24} &
    \multicolumn{1}{r|}{0/24}
    \\ \hline 
  \end{tabular}
  \end{small}
  \end{table*}
\subsection{Identifying Acceleration Opportunities}
\label{sec.compilation-stats}

%
We took the \AppNum DL applications, developed by different teams in different DSLs, and compiled them for the three target accelerators.
%
Our compiler successfully generated code that exploits the accelerators for supported computations.
%


Table~\ref{tab.compilation} shows the compilation statistics of using exact matching and flexible matching.
%
%
%
Note that some accelerator operators correspond to multiple Relay operators; in particular, 
the LSTM RNN in LSTM-WLM corresponds to 566 Relay operators 
and maps to \textit{one} FlexASR operator, which shows \TLA effectively overcoming a dramatic granularity mismatch between the compiler IR and accelerator operators.
%
%
%
%

Our results demonstrate \TLA's viability across a range of DL applications and accelerators with the successful identification of acceleration opportunities
%
and provide
evidence for the utility of flexible matching. 
%
%
%
For example, the linear layer rewrite 
(\S\ref{sec.method.flexible})
  resulted in 66 invocations of FlexASR's linear layer
  in Transformer and 38 in ResMLP, in comparison to exact matching that produced no match. 
Furthermore, certain Glenside rewrites~\cite{smith2021pure} that implement the \texttt{im2col} optimization~\cite{chellapilla2006high} 
rewrite 2D convolutions into matrix multiplications;
for VTA, this resulted in \emph{additional} 35 invocations in EfficientNet, 22 in ResNet-20, and 40 in MobileNet-V2.
Hence, flexible matching allowed us to support 2D convolutions on VTA even when 
there is no \mapping that maps 2D convolutions to VTA instructions.
Another rewrite
  that turns lone matrix multiplications
  into linear layers
  (by a zero-vector bias)
  works in concert with the \texttt{im2col} rewrites,
  resulting in offloads of 2D convolutions
  onto FlexASR  
  in EfficientNet, MobileNet-V2, and ResNet-20---%
  thus allowing an accelerator for NLP applications
  to also accelerate vision applications.
%
Note that these additional acceleration opportunities
  were identified automatically
  and are examples of \textit{emergent effects} resulting from simple, reusable (accelerator-agnostic) compiler IR rewrite rules.

We additionally
  evaluate the robustness
  of flexible matching
  by comparing
  the three implementations of ResNet-50
  from MLPerf~\cite{reddi2020mlperf}
  in Table~\ref{tab.compilation}, right.
Their Relay representations 
  differed in subtle ways (such as in reshaping operators)%
  \footnote{
For example, the TensorFlow implementation
  takes data in NHWC format rather than NCHW;
  Glenside can rewrite convolutions to use NCHW.}
  and are reflected in the difference in results of exact matching.
Flexible matching found the same (increased) number of matches for each accelerator, regardless of its source DSL.

\subsection{Per-Operator Evaluation}



Although evaluating individual operators
  does not suffice to characterize
  how an accelerator performs on a full application,
  it is a basic first step
  and provides 
  insights on the identified acceleration opportunities.
Here, we discuss functional validation and performance evaluation at the operator level.

\subsubsection{Functional Validation}

\begin{table}
\caption{
\textbf{
Simulation-based validation results for checking {\mapping}s (partial).
}
The average relative error (Avg. Err.) and the standard deviation (Std. Dev.) of errors are measured 
over 100 test inputs. 
For VTA, 
there was no error because the host supports 
8-bit integer operations.
}
\label{tab.layer-sim}
\centering
\begin{small}
\begin{tabular}{|c|l|l|r|r|}
\hline
  & Accel. & Operation & Avg.Err. & Std.Dev.  \\
  \hline \hline
  1 & VTA & All ops & 0.00\% & 0.00\% \\ 
  2 & HLSCNN & Conv2D & 1.78\% & 0.16\% \\ 
  3 & FlexASR & LinearLayer & 0.84\% & 0.29\% \\ 
  4 & FlexASR & LSTM & 1.21\% & 0.19\% \\ 
  5 & FlexASR & LayerNorm & 0.27\% & 0.20\% \\ 
  6 & FlexASR & MaxPool & 0.00\% & 0.0\% \\ 
  7 & FlexASR & MeanPool & 1.79\% & 0.28\% \\ 
  8 & FlexASR & Attention & 4.22\% & 0.09\% \\ 
\hline
\end{tabular}
\end{small}
\end{table}

The {\TLA} methodology readily enables operator-level validation through auto-generated ILA simulators.
In our experiments, we 
compared the outputs of the accelerator ILA simulator and those of TVM's runtime on host.
The accelerator ILA simulators precisely model the data types used by the accelerators.
%
%
For the reference results (TVM's runtime), we use 8-bit integer for comparing against VTA and 32-bit floating point for the other accelerators, as these are the closest host processor data types to those used by the accelerators. 
We measure the relative errors by using the standard Frobenius Norm~\cite{lapack} for the tensors based on the reference and accelerator generated output values as follows:
$Error = \| Out_{ref} - Out_{acc} \|_F / \| Out_{ref} \|_F$.

Table~\ref{tab.layer-sim} shows a selected subset of the validation results: four {\mapping}s (Rows~1-4) that are used in the full application compilation (Table~\ref{tab.compilation}) and four additional mappings for non-trivial operations (Rows~5-8).
%
%
%
Note that some mappings introduce no numerical differences; e.g., the TVM runtime supports 8-bit integer execution, so the results for VTA match perfectly.
%
For other mappings, we see deviations caused by the custom numerics, especially for complex operators such as the attention operator on FlexASR.  
Such deviations should be carefully assessed
  in the context of application-level validation,
  as even small deviations could accumulate and affect the final accuracy.


\subsubsection{Performance Evaluation}

We also evaluated the performance gain of offloading operations from the host to accelerators 
using cycle counts as the performance metric, since we did not have clock frequencies for an SoC containing the host and accelerators.
For accelerators, we derived the cycle counts based on their cycle-accurate models
(VTA's Chisel model
and FlexASR's and HLSCNN's SystemC models).
For the host, we measured averaged cycle counts (1000 random inputs) in TVM's runtime
on one pinned EPYC-7532 core.

\begin{figure}
  \centering
  \includegraphics[width=.55\linewidth]{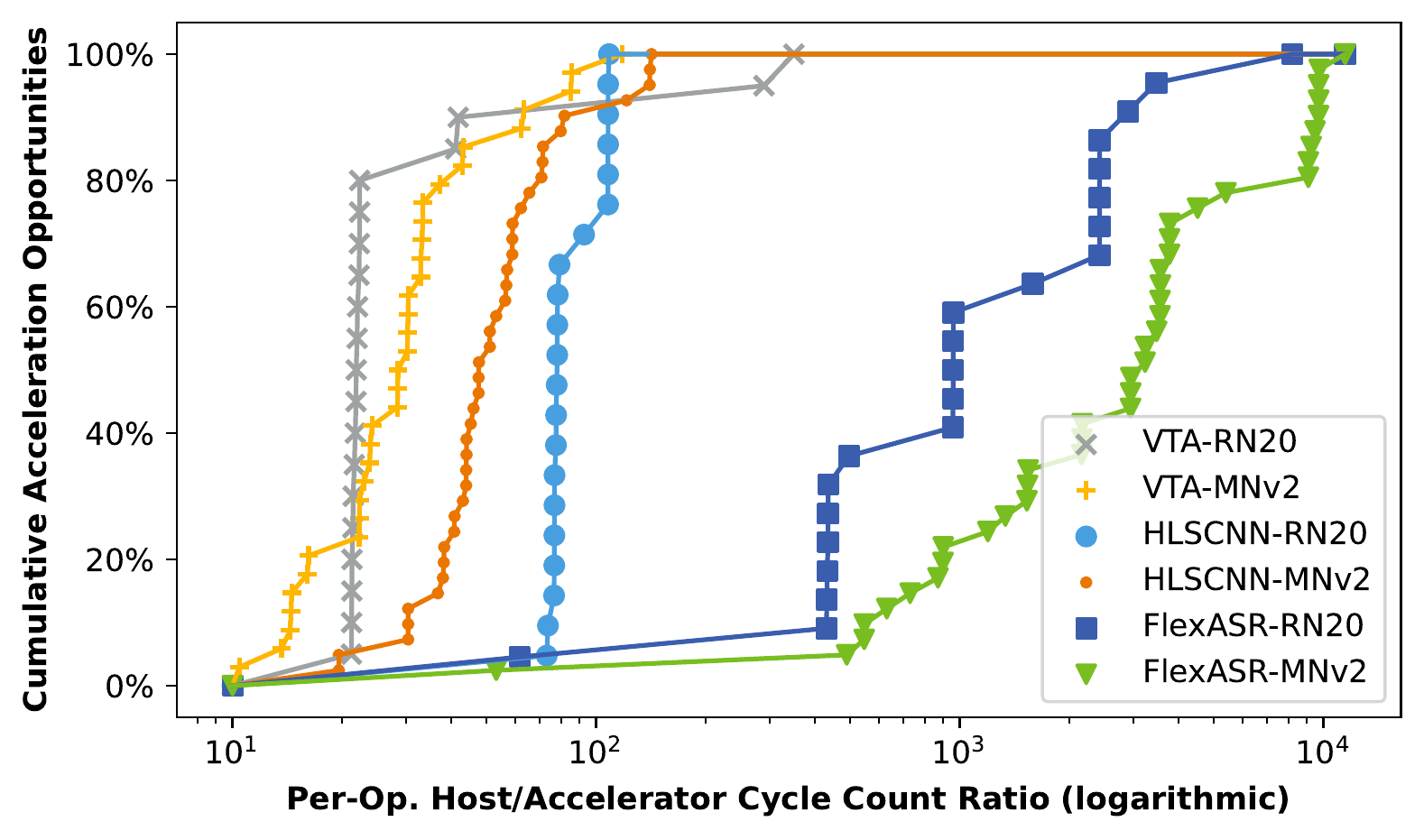}
  \caption{
\textbf{
Cumulative distribution of per-operator performance gains of all identified acceleration opportunities} in ResNet-20 (RN20) and MobileNet-V2 (MNv2) on the three accelerators.
Each point represents an operation offloaded from the host to the accelerator (as identified by flexible matching, Table~\ref{tab.compilation}).
The $x$-axis shows the host-to-accelerator cycle count ratio of each offloaded operation 
and the $y$-axis shows the cumulative distribution of offloaded operations. Points and plots more to the right are better; e.g., coarse-grained operators, supported with higher parallelism in FlexASR, offer greater speedup compared to the fine-grained operators in VTA. 
}
  \label{fig.performance}
\Description{}
\end{figure}

Fig.~\ref{fig.performance} shows the performance gains (ratio of host to accelerator cycles) of all identified acceleration opportunities in ResNet-20 and MobileNet-V2 when operations are offloaded from the host to VTA, HLSCNN, and FlexASR, respectively. 
Overall, as expected, all offloads resulted in performance gains relative to the host; we also see that accelerators providing coarser-grained operators (e.g., FlexASR), supported with higher parallelism, achieve higher performance gain per operator compared to finer-grained accelerators like VTA. 
\subsection{Application-Level Validation Through Co-Simulation}
\label{sec.end-to-end}

We  performed application-level co-simulation
  by using the
  ILAng-generated simulators for accelerator computations
  and the host CPU for the rest of the computation.
%
%
We considered three applications,
  which between them provide opportunities to use
  each of the three accelerators: 
\begin{inlinelist}
  \item LSTM-WLM, where we accelerate linear layer and LSTM operations on FlexASR;
  \item ResNet-20, where we accelerate convolutions on HLSCNN and linear layers on FlexASR; and
  \item MobileNet-V2, where we accelerate convolutions and linear layers as in ResNet-20 and additionally accelerate both these operations on VTA (due to the \texttt{im2col} rewrites).
\end{inlinelist}
In ResNet-20 and MobileNet-V2,
  we were able to
  \emph{explore using HLSCNN and FlexASR together and separately}, simply by varying which 
  {\mapping}s
  we included in flexible matching.

We trained and validated the LSTM-WLM model using the WikiText-2 dataset~\cite{merity2016pointer}.
The image classification models (MobileNet-V2 and ResNet-20) were trained and validated using the CIFAR-10 dataset~\cite{cifar10}.

Table~\ref{tab.verif-sim} shows the application-level co-simulation results.
%
%
%
%
%
%
For LSTM-WLM,
  the application-level results using the accelerators
  did not differ greatly
  from the reference results.
In the case of FlexASR,
  this was the \textit{first time}
  it had been run end-to-end on a full application---%
  this provided validation for its AdaptivFloat data type.
For VTA on MobileNet-V2,
  there was a small decrease in accuracy
  that may be attributed
  to quantization error.%
\footnote{We apply a form of uniform quantization~\cite{jacob2017quantization},
  which involves scaling the results
  based on the floating point reference results.}

\begin{table*}
  \caption{
  \textbf{Application-level co-simulation results.}
  In each test, we evaluated 2000 CIFAR-10 images (for vision tasks) or 100 WikiText-2 sentences (for text generation) that were evenly sampled from the corresponding dataset.
  The reference results were obtained by running tasks in the original frameworks (MxNet for ResNet-20, PyTorch for the rest).
  The original results are 
  for the initial accelerator designs, modeled in ILA.
  The updated results, where provided, were obtained by modifying the ILA specifications
  according to 
  design revisions suggested by the accelerator developers.
  We measured the accuracy for image classification tasks (ResNet-20, MobileNet-V2) and perplexity for text generation (LSTM-WLM).}
  \label{tab.verif-sim}
  \centering
  \begin{small}
  \begin{tabular}{|l|c|c|c|c|r|}
  \hline
  \multicolumn{1}{|c|}{Application} &
    \multicolumn{1}{c|}{Processing Platform} &
    \multicolumn{1}{c|}{Reference Result$^\ast$} &
    \multicolumn{1}{c|}{Original Result} &
    \multicolumn{1}{c|}{Updated Result} &
    \multicolumn{1}{c|}{Avg. Sim. Time$^\dagger$} \\
    \hline \hline

  LSTM-WLM & 
    FlexASR & 
    122.15 & 
    121.97 & 
    N/A &
    22.4s \\
    \hline


  ResNet-20 & 
    FlexASR &
    91.55\% &
    91.50\% &
    N/A &
    11.6s \\
    
   &
    HLSCNN &
    91.55\% &
    \cellcolor[HTML]{E9CECE}29.75\% &
    \cellcolor[HTML]{DDEFDE}92.10\% &
    7min 3s \\
    
   &
    FlexASR \& HLSCNN & 
    91.55\% & 
    \cellcolor[HTML]{E9CECE}29.15\% & 
    \cellcolor[HTML]{DDEFDE}91.85\% & 
    7min 6s \\ 
    \hline

  MobileNet-V2 &
    VTA &
    92.40\% &
    89.40\% &
    N/A &
    20min 15s \\
  
   &
    FlexASR &
    92.40\% &
    92.30\% &
    N/A &
    18.1s \\

   & 
    HLSCNN & 
    92.40\% & 
    \cellcolor[HTML]{E9CECE}10.35\% & 
    \cellcolor[HTML]{DDEFDE}91.50\% & 
    20min 33s \\
    
   & 
    FlexASR \& HLSCNN & 
    92.40\% & 
    \cellcolor[HTML]{E9CECE}10.35\% & 
    \cellcolor[HTML]{DDEFDE}91.20\% & 
    21min 01s \\


    \hline
  \end{tabular}
  \end{small}
  \begin{tablenotes}
    \item $\ast$ The reference result does not represent the best achievable accuracy/perplexity of the model on the given dataset. This table is intended for comparing the application-level results on different processing platforms.
    \item $\dagger$ Average simulation time of running one data point (e.g., an image or a sentence) on an AMD EPYC-7532 core.
  \end{tablenotes}
\end{table*}

However, the initial results for ResNet-20 and MobileNet-V2
  using HLSCNN
  revealed a large loss in accuracy.
We noticed that the linear layers 
  accelerated by FlexASR 
  did not impact the final accuracy,
  suggesting the issue stemmed from HLSCNN
  (for which this was also the first time it was run in an end-to-end application).
We then instrumented
  our {\TLA} prototype 
  to record additional information
  for each accelerator invocation,
  such as input and output ranges.
This helped 
  the accelerator developers
  determine that the loss of accuracy
  was due to a lack of dynamic range in the data type:
  weight data values 
  in HLSCNN's 2D convolutional layers
  were heavily quantized
  due to the narrow value range
  of their 8-bit fixed point representation.
After we updated the ILA specification (a much easier task than modifying the RTL implementation) based on the developers' suggestion to expand the fixed point representation to 16 bits and adjust the binary points in inputs' and accumulators' fixed point data types, the 
accuracy recovered.
  This case study readily demonstrates
  how the {\TLA} methodology
  \textit{facilitates debugging and improving accelerator designs
  with rapid turnaround.}

The overall results in Table~\ref{tab.verif-sim} reaffirm the need for application-level validation, especially for accelerators utilizing custom numerics.
%
%
%
Thanks to formal ILA models, \TLA provides quick design space exploration and numerics tuning without hardware engineering overhead 
in each design iteration.
%
%
Further, it provides handy debugging information and efficient simulation---%
for FlexASR, the ILA simulator yields a 30$\times$ speedup on average compared to RTL simulation.
\subsection{System Deployment and FPGA Emulation}
\label{sec.eval-fpga}

As an additional demonstration of {\TLA}, we explored its use in 
  compiling workloads
  to a real hardware platform.
  Specifically, 
  we used our prototype to compile workloads
  to an FPGA emulation of FlexASR.%
\footnote{We synthesized and placed-and-routed the FlexASR accelerator on a Xilinx Zynq ZCU102 FPGA, which consumed 86\% of the available LUT resources.
Due to the significant engineering overhead of FPGA emulation, FlexASR is the only accelerator we deployed on an FPGA.}
%
We configured our prototype to lower
 FlexASR ILA instructions
 to the corresponding MMIO commands for FlexASR, 
 passing them to the FPGA using the Xilinx SDK~\cite{xsdk}.
%
Next, we compiled and executed synthetic workloads in which LSTM layers and linear layers were offloaded to the FlexASR accelerator.
The results matched those of the ILAng-generated simulator bit for bit, providing validation for 
the custom numerics.
This is a proof of concept for utilizing the {\TLA} methodology for an actual deployment, above and beyond simulation-based testing.
%


\section{Related Work}
\label{sec.related}

\paragraph{Software/Hardware Co-Design}

Recent work on accelerator generation and integration~\cite{
    bahr2020creating, truong2020fault}
  has explored adding support in the Halide~\cite{ragan2013halide}
  compiler flow for specialized Coarse-Grained Reconfigurable Array (CGRA) accelerators.
That work composes an
  impressive array of custom tools to
  generate and verify specialized CGRA accelerators
  and also map Halide program fragments
  down to accelerator invocations.
HeteroCL~\cite{lai2019heterocl} also provides
  a similar custom flow.
By contrast, the \TLA methodology supports 
  software/hardware co-design by mitigating impedance mismatches
  between the granularity of high-level 
  DSLs and \textit{near-arbitrary} accelerators;
  because of the flexibility of the ILA,
  the \TLA methodology is applicable to a
  broader class of compilers and accelerators.

\paragraph{Pattern Matching Accelerator Calls}


The most closely related work to flexible matching is from 
\begin{inlinelist}
  \item TVM BYOC~\cite{chen2021byoc}, which only provides exact syntactic matching as discussed in \S\ref{sec.background}, and
  \item Glenside~\cite{smith2021pure}, which, prior to this work, had not been integrated into a compilation pipeline nor used to target custom accelerators.
\end{inlinelist}
Past work has also explored rewrite-based techniques for
  automatically inferring instruction selection passes
  between ISAs~\cite{
    ramsey2011resourceable,
    dias2010automatically}
  and in the context of superoptimization~\cite{
    bonsal-so,
    bonsal-so-translate}.
Rewriting in \TLA instead operates on a high-level IR
  to expose opportunities to invoke code generators,
  rather than performing low-level code generation directly.
Equality saturation has been
  used in the context of
  ML and DSP compilers for
  optimization~\cite{
    yang2021equality,
    alexa-dsp-eqsat,
    caviar-cc22}.
There has also been significant work on
  ML and HPC compiler frameworks with
  varying degrees of support for
  targeting custom accelerators~\cite{
    ragan2013halide,
    AtlPopl22,
    chen2018tvm,
    moreau2019hardware,
    lattner2021mlir}.
To the best of our knowledge,
  none of these frameworks provides support for
  testing prototype accelerators
  designs end-to-end on
  unmodified source applications.

\paragraph{Validating and Verifying Accelerator Calls}


Tools like Verilator~\cite{verilator} and Cuttlesim~\cite{pitclaudel2021cuttlesim} enable cycle-accurate RTL simulation,
but are too slow to enable application-level co-simulation. Co-simulation using faster high-level SystemC~\cite{SystemC} models partially address this gap; however, the SystemC models need to be independently written and, unlike ILA models, do not have a clear formal verification path to RTL. Further, 
general SystemC models do not target MMIO interfaces and may have arbitrary levels of detail. 
Other work has targeted formal verification of code generation for accelerators~\cite{AtlPopl22,ExoPldi22}, but does not have a path to RTL design verification that is possible with ILAs. 
%

\paragraph{Support for Diverse DSLs}

Our \TLA prototype supports importing from various DL frameworks (including MxNet~\cite{chen2015mxnet}, PyTorch~\cite{paszke2019pytorch}, TensorFlow~\cite{abadi2016tensorflow}, ONNX~\cite{linux2019onnx}, and CoreML~\cite{apple2022coreml}) via TVM's importers to the Relay IR.
Similar to Relay, MLIR~\cite{lattner2021mlir} now also  supports importing models from 
various frameworks; the \TLA approach complements such flows as demonstrated in our prototype.
In contrast, Exo~\cite{ExoPldi22} presently has no support for importing other representations.

\section{Conclusions}
\label{sec.conclusion}


In this work we address the key gaps 
hindering
application-level evaluation of accelerator designs, especially during early design stages.
We propose the \TLA methodology that contributes 
\begin{inlinelist}
\item the use of a formal software/hardware interface for specifying accelerator operations,
which enables identifying acceleration opportunities and automatically generating correct high-level simulators, and 
\item compiler rewrites and equality saturation for flexible matching, which 
facilitates automatically searching through a large space of equivalent programs to find 
acceleration opportunities. 
\end{inlinelist}
We provide a \TLA prototype implementation for DL applications using the TVM 
and ILAng frameworks
and evaluate it 
through automated compilation of \AppNum applications on three different accelerator platforms. 

\bibliographystyle{ACM-Reference-Format}
\bibliography{references}

\end{document}